\documentclass[10pt]{article}
\usepackage{amsmath,amssymb,amsfonts}
\usepackage{units}
\usepackage{bbold,euler}%opcoes de fonte: mathptmx, helvet, eulervm, avant, fourier, newcent, mathpazo,euler,newcent,sansmath
\usepackage{graphicx}% Include figure files
\usepackage{dcolumn}% Align table columns on decimal point
\usepackage{bm,bbm,mathtools,esvect}
\usepackage{slashed}
\usepackage{esvect}
\usepackage{fullpage}
\usepackage[utf8]{inputenc}
\title{\bf EXPECTATION VALUE DYNAMICS WITHIN REAL HILBERT SPACE QUANTUM MECHANICS}

\author{{\sf SERGIO GIARDINO\footnote{\tt sergio.giardino@ufrgs.br}}\\
\\
\small \it Departamento de Matem\'atica Pura e Aplicada \\
\small \it Universidade Federal do Rio Grande do Sul (UFRGS)\\
\small \it Caixa Postal 15080, 91501-970  Porto Alegre RS \\
\small \it Brazil}
\begin{document}
\date{} % Remove a data
\maketitle

\begin{abstract}
\noindent Dynamic equations concerning physical expectation values have been examined in terms of the real Hilbert space approach to quantum mechanics. The considered cases involve complex wave functions, as well as quaternionic wave functions. The consistency of the formalism has been verified in terms of the continuity equation, the classical limit, and generalizations of the quantum Lorentz force, and the Virial theorem. Besides testing the consistency of the real Hilbert space approach, generalized position and angular momentum operators have been introduced, and inspire exciting directions for further research.

\vspace{2mm}

\noindent keywords: quantum mechanics; formalism; other topics in mathematical methods in physics

\vspace{1mm}

\noindent pacs numbers: 03.65.-w; 03.65.Ca; 02.90.+p.
\end{abstract}

\vspace{1cm}

%\pagebreak

\hrule
\tableofcontents
%{\parskip - 0.0mm \tiny{\tableofcontents}}
\vspace{1cm}
\hrule
\pagebreak
%%%%%%%%%%%%%%%%%%%%%%%%%%%%%%%%%%%%%%%%%%%%%%%%%%%%%%%%%%%%%%%%%%%%%%%%%%
%%%%%%%%%%%%%%%%%%%%%%%%%%%%%%%%%%%%%%%%%%%%%%%%%%%%%%%%%%%%%%%
\section{ INTRODUCTION\label{OI}}
%%%%%%%%%%%%%%%%%%%%%%%%%%%%%%%%%%%%%%%%%%%%%%%%%%%%%%%%%%%%%%%
%%%%%%%%%%%%%%%%%%%%%%%%%%%%%%%%%%%%%%%%%%%%%%%%%%%%%%%%%%%%%%%%%%%%%%%%%%

The generality of a physical theory determines which problems the very same theory may solve. Nevertheless, the theme of the theoretical generality may arise from various sources. Unsolved problems of a particular theory may indicate a theoretical insufficiency  requiring a different physical model, or even a more robust mathematical framework for the theory. Unexplained experimental data are of course the most important sort unsolved questions, but a purely theoretical issue like the quantization of general relativity can also be considered an unsolved problem within a physical theory. The resolution of these particular topics requires some intervention, whose result would hopefully be a more general theory in whose ambit the open questions are naturally solvable. 

The discussion of generality in quantum mechanics rose up quite early because of the complex character of the Schrödinger equation, as demonstrates the Birkoff and von Neumann's article of 1936 \cite{Birkhoff:2017kpl}. On that paper, the investigation of a quantum logic independent of the number field engendered the hypothesis of a quantum theory expressed in terms of non-complex numbers, namely the reals, the quaternions, or the octonions. As a further hypothesis within this background, this article is part of the effort to understand the generality of the mathematical structure of quantum mechanics, hoping that unexplained physical questions could be addressed in terms of a more powerful theoretical framework. The real Hilbert space approach used here in some sense unifies complex quantum mechanics ($\mathbbm C$QM), and quaternionic quantum mechanics ($\mathbbm H$QM) by enabling the real Hilbert space to admit quaternionic wave functions, a formalism that also removes the hermiticity condition that is usual in the standard complex quantum mechanics ($\mathbbm C$QM), and consequently generates a more general and less constrained theory. The real Hilbert space formalism emerges as a solution to the breakdown of the classical limit within the anti-hermitian construction of $\mathbbm H$QM ({\em cf.} Sec. 4.4 of \cite{Adler:1995qqm}), catalyzing the introduction of the real Hilbert space formalism \cite{Giardino:2018lem,Giardino:2018rhs}, suitable for several non-relativistic quaternionic models like the geometric phase \cite{Giardino:2016xap}, the free particle \cite{Giardino:2017yke,Giardino:2017pqq,Giardino:2024tvp}, the Virial's theorem \cite{Giardino:2019xwm}, the rectangular potential \cite{Giardino:2020cee}, the quantum  scattering \cite{Giardino:2020ztf,Hasan:2019ipt}, the harmonic oscillator \cite{Giardino:2021ofo},  spin \cite{Giardino:2023spz}, and the generalized wave equation \cite{Giardino:2023uzp}. Moreover, the real Hilbert space also suits well also to relativistic models \cite{Giardino:2021lov,Giardino:2021mjj}, including field quantization \cite{Giardino:2022kxk,Giardino:2022gqn}.

On the other hand, the interpretation of the quaternionic wave functions as suitable for the expression of self-interactions \cite{Giardino:2024tvp} invites us to revisit several results of $\mathbbm H$QM obtained before introducing the self-interacting hypothesis.  In fact, the autonomous particle considered in \cite{Giardino:2024tvp} is an important physical motivation to this article, and of course could be also an example of the formalism considered here. Hence, in the present article one recalls a former result concerning the dynamics of expectation values \cite{Giardino:2018rhs}, and the generalized linear momentum \cite{Giardino:2019xwm} within an unified theoretical template capable of expressing complex and quaternionic wave equations, as well as the respective expectation value dynamics, quantum Lorentz force, and the Virial theorem. The real Hilbert space approach permits to deep the understanding of the previous results, unraveling the self-interacting role of complex and quaternionic components of the wave functions and quantum operators, and hence generating corrections to the expectation values.

One must also acknowledge this article as part of a discussion concerning  the hypothesis of a real quantum mechanics ($\mathbbm R$QM), which includes favorable considerations \cite{Finkelstei:2022rqm,Chiribella:2022dgr,Zhu:2020iml,Fuchs:2022rih,Vedral:2023pij}, as well as arguments against their viability \cite{Renou:2021dvp,Chen:2021ril,Wu:2022vvi,Li:2021uof,Batle:2024oqg}. However, the real Hilbert space presented here does not require real wave functions, and can be considered thus an alternative that could unify these formalisms within a single theory with generalized wave functions. Finally, additional arguments involving of the complex character of quantum mechanics also include quantization methods \cite{Prvanovic:2020clx}, mathematical formulation \cite{Takatsuka:2025piu,Volovich:2025rmi,Hita:2025okv}, imaginarity \cite{Gour:2018qiq,Xu:2023xdb,Wu:2023lyi,Zhang:2025jdb,Saruhan:2024ifh},
locality \cite{Feng:2025eci}, unitarity \cite{Fernandes:2024sik}, and complex potentials \cite{Khantoul:2022bam,Cannata:2012uc}, but one expects that all these topics of research may be explained after the understanding of the generality question, and the results contained in this article are intended to be an additional step in this direction.

The results of the article are organized within three sections, comprising one complex case, and two quaternionic cases. One remarks all these sections have exactly the same subsections, and this is done in order to facilitate the comparison of the cases. The complex case is of course the simplest, and constitutes the model to be followed in the more involved quaternionic cases.

%%%%%%%%%%%%%%%%%%%%%%%%%%%%%%%%%%%%%%%%%%%%%%%%%%%%%%%%%%%%%%%%%%%%%%%%%%
%%%%%%%%%%%%%%%%%%%%%%%%%%%%%%%%%%%%%%%%%%%%%%%%%%%%%%%%%%%%%%%
\section{COMPLEX SOLUTION \label{CSO}}
%%%%%%%%%%%%%%%%%%%%%%%%%%%%%%%%%%%%%%%%%%%%%%%%%%%%%%%%%%%%%%%
%%%%%%%%%%%%%%%%%%%%%%%%%%%%%%%%%%%%%%%%%%%%%%%%%%%%%%%%%%%%%%%%%%%%%%%%%%

This section is intended to gather the complex results that will be extended to the two further quaternionic cases. Additionally, in this section one generalizes to the real Hilbert space the usual complex quantum mechanics ($\mathbbm C$QM) defined within the complex Hilbert space, and  this is done using the complex extensions of quantum operators that only make sense after introducing the real Hilbert space.

\paragraph{COMPLEX DYNAMICS} One thus starts recalling the motion of a quantum particle of mass $m$ codified in terms of a complex wave function $\psi$ that  satisfies the Schrödinger equation,
\begin{equation}\label{cso01}
	i\hbar\frac{\partial \psi}{\partial t}=\widehat H\psi,
\end{equation}
where the Hamiltonian operator $\widehat H$ conforms to
\begin{equation}\label{cso11}
	\widehat H=\frac{1}{2m}\widehat \Pi^2+U,
\end{equation}
comprising $U$ as the scalar potential, and the $\widehat{\bm \Pi}$ as the generalized linear momentum operator
\begin{equation}\label{cso04}
	\widehat{\bm\Pi}=\widehat{\bm p}-\frac{q}{c}\bm{\mathcal  A},
\end{equation}
where to $\bm{\mathcal A}$ corresponds the potential vector, and to $q$ and $c$ respectively correspond the electric charge, and the speed of the light in vacuum. Of course, 
\begin{equation}
	\widehat{\bm p}=-i\hbar\bm\nabla,
\end{equation}
is the usual quantum linear momentum operator, and to $\hbar$ corresponds the Planck's constant.
The above elements can be immediately generalized admitting a complex scalar potential $U$, as well as a complex vector potential $\bm{\mathcal A}$. However, this situation imposes the breakdown of the hermiticity condition, and therefore neither
\begin{equation}\label{cso20}
	\widehat H\neq\widehat H^\dagger,\qquad\mbox{nor}\qquad \widehat{\bm\Pi}\neq\widehat{\bm\Pi}^\dagger,
\end{equation}
where of course $\widehat H^\dagger$ and $\widehat{\bm\Pi}^\dagger$ are hermitian conjugates. A consistent quantum theory incorporating such non-hermitian operators can be achieved  after imposing the expectation value of an arbitrary operator $\widehat{\mathcal O}$ to assume
\begin{equation}\label{cso02}
	\big<\mathcal O\,\big>=\int\Big\{\psi,\,\widehat{\mathcal O}\psi\Big\} d\bm x,
\end{equation}
where the curly bracket term corresponds to the probability density defined as
\begin{equation}\label{cso05}
\Big\{\psi,\,\widehat{\mathcal O}\psi\Big\}=\frac{1}{2}\left[\psi^\dagger\widehat{\mathcal O}\psi+\psi\big(\widehat{\mathcal O}\psi\big)^\dagger\right].
\end{equation}
The expectation value (\ref{cso02}) and the probability density (\ref{cso05}) comprise the core structure of the real Hilbert space, generating  real quantities irrespective the character of the  operator $\widehat{\mathcal O}$, and therefore hermitian operators are no longer required for real expectation values. Immediately, the symbol $z^\dagger$ can be simply identified to the complex conjugate of $z$.

An immediate consequence of the real expectation value (\ref{cso02}) comprises that the complex component of $\bm{\mathcal A}$ does not contributes to the expectation value of $\big<\bm\Pi\big>$, but it will seen in the discussion of the continuity equation as related to non-conservative phenomena.  The complex generalization can be also extended to a generalized position operator
\begin{equation}\label{cso06}
	\widehat{\bm z}=\widehat{\bm r}+\widehat{\bm s} i
\end{equation}
in terms of the ordinary three-dimensional position vector
\begin{equation}
	\widehat{\bm r}=\big(x,\,y,\,z\big)
\end{equation}
and a novel vector real function
\begin{equation}
	\widehat{\bm s}=\bm s(\bm r).
\end{equation}
One may initially suppose the $\widehat{\bm z}$ operator as speculative and arbitrary, however one observes the interesting physical property that the imaginary component does not change the position of the particle, or explicitly
\begin{equation}\label{cso21}
	\big<\bm z\big>=\big<\bm r\big>,
\end{equation}
as can be immediately verified after substituting (\ref{cso06}) in (\ref{cso02}). Therefore, one may understand the imaginary contribution in the same fashion as a gauge potential that does not change physical observation,  and consequently the position operator requires the imaginary component in order to provide a complete description of the physical system. Although the expectation value does not change, the imaginary component addresses concrete consequences in terms of the commutator relation, whose expression in component notation reads
\begin{equation}\label{cso22}
\Big[\widehat z_k,\,\widehat p_\ell\big]=\hbar\Big(i\delta_{k\ell}-\partial_\ell s_k\Big),
\end{equation}
where  $\partial_\ell$ of course represents the partial derivative with respect to the $x^\ell$ coordinate. The $\bm s\to\bm 0$ of course recovers the usual commutation relation, as expected. Although (\ref{cso22}) is a novelty, this kind of commutator has already been reported \cite{Giardino:2025jni}, where the presence of imaginary and real components on the right hand side indicates the existence of non-stationary phenomena, a physical interpretation to be further discussed hold along this paper.  Moreover, the commutator also finds relation to the generalized uncertainty principle \cite{Bosso:2023aht}, thus being a natural way to obtain more general commutation relations in order to reach a wider rank of physical phenomena.

As a final comment, it has been shown \cite{Giardino:2018rhs,Giardino:2019xwm} that Fourier expansions in terms of real coefficients are determined using (\ref{cso02}), consequently establishing that quaternionic functions may constitute the basis of a real Hilbert space. Further essential tests of the real Hilbert space theory will be conducted in this section in terms of complex wave functions, namely the continuity equation, the classical limit, the expectation value dynamics, the Lorentz force, and the Virial theorem. They will constitute the template for testing the quaternionic wave functions in the remaining sections of the paper.

\paragraph{CONTINUITY EQUATION} Multiplying the wave equation (\ref{cso01}) by $i\psi^\dagger$ and taking the real part, one obtains the continuity equation
\begin{equation}\label{cso03}
	\frac{\partial \rho}{\partial t}+\Big(\bm\nabla+\bm\gamma\Big)\bm{\cdot J}=g,
\end{equation}
where the probability density $\rho$ and the probability density current $\bm J$ conform to
\begin{equation}
	\rho=\psi\psi^\dagger,\qquad\qquad
	\bm J=\frac{1}{2m}\left[\psi^\dagger\Big(\widehat \Pi\psi\Big)+\psi\Big(\widehat \Pi\psi\Big)^\dagger\right].
\end{equation}
One observes two non-conservative quantities in (\ref{cso03}), respectively
\begin{equation}
	\bm \gamma=i\frac{\bm {\mathcal A}^\dagger-\bm{\mathcal A}}{\hbar},\qquad g=i\frac{U^\dagger-U}{\hbar}\rho.
\end{equation}
Both of these terms are generated by imaginary components of $U$ and $\bm{\mathcal A}$, and thus recovering the usual $\mathbbm C$QM result in case of real $U$ and $\bm{\mathcal A}$. Non-conservative terms of the continuity equation are associated to non-stationary physical phenomena, like inelastic scattering, and dissipative processes ({\it cf.} \cite{Schiff:1968qmq} Section 20). Consequently, the continuity equation provides the physical interpretation of the complex components of both of the potential terms, and one remarks the presence of the $\bm\gamma$ term in (\ref{cso03}) to be a novel result as a consequence of the complex vector potential within the generalized linear momentum operator (\ref{cso04}). This interpretation of the imaginary terms of both of the potential terms is also coherent to the imaginary component of the generalized position operator $\widehat{\bm z}$ discussed in (\ref{cso22}).

\paragraph{CLASSICAL LIMIT} Taking advantage of the generalized position operator, and of the continuity equation, one obtains the generalized time evolution
\begin{equation}\label{cso07}
	\frac{d\big< \bm z\big>}{dt}=\frac{1}{m}\big<\bm \Pi\big>+\Big<\big( g-\bm{\gamma\cdot J}\big)\bm z\Big>.
\end{equation}
Because $\,g-\bm{\gamma\cdot J}\,$ is a real quantity, the imaginary component of $\widehat{\bm z}$ does not contribute to the expectation value, and therefore the generalized position operator  (\ref{cso06}) does not contribute to the expectation value. If the non-conservative contributions to the continuity equation are set to zero, one recovers the ordinary $\mathbbm C$QM result, as expected.  One may rise the question whether it is possible to remove the second term on the right hand side of (\ref{cso07}) in the same fashion of the `renormalization' scheme  done in \cite{Sergil:2012nho}. In principle, this target cannot be reached because the components $g$ and $\gamma$ cannot be removed from the continuity equation (\ref{cso03}) because of their non-conservative physical content. However, the precise relation between the results of this article and those of \cite{Sergil:2012nho} is certainly an interesting direction for future research.

After observing in (\ref{cso07}) that the classical relation between velocity and momentum also holds for quantum operators, one acts in accordance to ordinary textbooks to obtain the relation between force and potential, namely
\begin{equation}\label{cso08}
	\frac{d\big< \bm p\big>}{dt}=-\left<\bm\nabla\left(\frac{U+U^\dagger}{2}\right)\right>+\left<\frac{U-U^\dagger}{i\hbar}\bm p\right>.
\end{equation}
Again, the classical relation between a conservative force and the gradient of a scalar potential is recovered for real scalar potentials, as desired. The above relations (\ref{cso07}-\ref{cso08}) do not hold within the complex Hilbert space formulation of quantum mechanics, where the imaginary component of the potential cannot be eliminated. 

%Moreover, qualitatively examining (\ref{cso07}-\ref{cso08}) as differential equations, one observes the imaginary component of the potentials as generating typically dissipative differential equations, also reinforcing the interpretation of these imaginary terms as corresponding to non-conservative physical processes.

\paragraph{EXPECTATION VALUES DYNAMICS} The well-known time evolution of quantum expectation value in the case of hermitian Hamiltonian operators simply reads
\begin{equation}\label{cso09}
\frac{d\big<\mathcal O\big>}{dt}=\left<\frac{1}{i\hbar}\big[\mathcal  O,\,H\big]\right>+\left<\frac{\partial\mathcal O}{\partial t}\right>+\frac{\partial\big<\mathcal O\big>}{\partial t}. 
\end{equation}
However, one can argument that the above equation also holds in case of non-hermitian Hamiltonian operators because the identity
\begin{equation}
	\widehat{H}\big(\widehat{\mathcal O}\psi\big)=i\hbar\frac{\partial}{\partial t}\big(\widehat{\mathcal O}\psi\big)=i\hbar\left(\frac{\partial \widehat{\mathcal O}}{\partial t}\psi+\widehat{\mathcal O}\frac{\partial\psi}{\partial t}\right),
\end{equation}	
and the wave equation (\ref{cso01}) jointly determine
\begin{equation}\label{cso10}
	\left<\frac{1}{i\hbar}\big[\mathcal  O,\,H\big]\right>+\left<\frac{\partial\mathcal O}{\partial t}\right>=0.
\end{equation}
Relation (\ref{cso10}) holds for non-hermitian operators, irrespective of the Hilbert space, but one should remember that only within the real Hilbert space that (\ref{cso10}) involve real quantities. Comparing (\ref{cso09}) and (\ref{cso10}), one thus argues in physical terms that equation (\ref{cso09})  accordingly holds for non-hermitian operators acting within a real Hilbert space. A similar generalization holds for quaternionic cases, as will be seen.

\paragraph{LORENTZ FORCE}  Using $\widehat{\mathcal O}=\widehat{\bm\Pi}$ and the Hamiltonian operator (\ref{cso11}) in (\ref{cso09}) permits to determine the dynamics of the generalized momentum (\ref{cso04}), in analogy to (\ref{cso08}). First of all,  considering the time derivatives on the right hand side of (\ref{cso09}) to disappear, one calculates
\begin{equation}\label{cso12}
	\Big[\widehat{\bm\Pi},\,\widehat H\Big]=\frac{1}{2m}\Big[\widehat{\bm\Pi},\,\widehat{\Pi}^2\Big]+\Big[\widehat{\bm\Pi},\,U\Big],
\end{equation}
and immediately obtaining
\begin{equation}
	\Big[\widehat{\bm\Pi},\,U\Big]=-i\hbar\bm\nabla U.
\end{equation}
The first commutator of the right hand side of (\ref{cso12}) can be decomposed in terms of its components as 
\begin{equation}
	\Big[\widehat{\Pi}_i,\,\widehat{\Pi}_j^2+\widehat{\Pi}_k^2\Big]=\widehat\Pi_j\Big[\widehat{\Pi}_i,\,\widehat{\Pi}_j\Big]+\Big[\widehat{\Pi}_i,\,\widehat{\Pi}_j\Big]\widehat \Pi_j+\widehat\Pi_k\Big[\widehat{\Pi}_i,\,\widehat{\Pi}_k\Big]+\Big[\widehat{\Pi}_i,\,\widehat{\Pi}_k\Big]\widehat \Pi_k,
\end{equation}
where $i,\,j$ and $k$ can assume non-repeated values of $1,\,2$ and $3$. Moreover, taking benefit of
\begin{equation}
	\Big[\widehat{\Pi}_i,\,\widehat{\Pi}_j\Big]=\frac{q}{c}\Big(\widehat p_j\mathcal A_i-\widehat p_i\mathcal A_j\Big)=i\hbar\frac{q}{c}\epsilon_{ijk}\big(\bm{\nabla\times{ \mathcal A}}\big)_k,
\end{equation}
where $\epsilon_{ijk}$ is the Levi-Cività anti-symmetric symbol, one defines the complex magnetic field
\begin{equation}
	\bm{\mathcal B}=\bm{\nabla\times{ \mathcal A}},
\end{equation}
and the expectation value dynamics thus formally reproduces the $\mathbbm C$QM result
\begin{equation}\label{cso13}
	\frac{d\big<\bm\Pi\big>}{dt}=\left<\frac{1}{i\hbar}\frac{ \bm{\widehat\Pi\times\widehat{\mathcal B}} - \bm{\widehat{\mathcal B}\times\widehat\Pi}}{2m}\right>-\big<\bm\nabla U\big>,
\end{equation}
that of course recovers the ordinary $\mathbbm C$QM outcome in case of real potentials. One has to point out the imaginary component of the real scalar potential $U$ not to contribute to the expectation value, but the vector potential contributes to it, as one considers that
\begin{equation}
\widehat{\bm\Pi}=\widehat{\bm\Pi}_0+\frac{q}{c}\bm A_1i,\qquad\bm{\mathcal B}=\bm B_0+\bm B_1i
\end{equation}
where 
\begin{equation}
	\widehat{\bm\Pi}_0=\widehat{\bm p}+\frac{q}{c}\bm A_0
\end{equation}
and $\bm B_0,\,\bm B_1,\,\bm A_0$ and $\bm A_1$ are real vectors, and after some manipulation one obtains (\ref{cso13}) to be
\begin{equation}\label{cso24}
	\frac{d\big<\bm\Pi\big>}{dt}=\left<\frac{1}{i\hbar}\frac{ \bm{\widehat\Pi}_0\bm{\times\widehat{\mathcal B}} - \bm{\widehat{\mathcal B}\times\widehat\Pi}_0}{2m}\right>-\big<\bm\nabla U\big>.
\end{equation}
Therefore, the $\bm A_1$ component of the potential vector $\bm{\mathcal A}$  contributes to the expectation value via the complex magnetic field $\bm{\mathcal B}$, only. Investigating the physical system that concretely admits such kind of description is an exciting direction of future research.

\paragraph{VIRIAL THEOREM}  A final consistency test of the real Hilbert approach to complex quantum mechanics considers the quantity $\bm{r\cdot p}$ to be invariant for periodic physical systems. Therefore, the quantum dynamics of expectation values requires that after setting 
\begin{equation}\label{cso23}
\widehat{\mathcal O}\to \widehat{\bm r}\bm\cdot\widehat{\bm p}
\end{equation} 
in (\ref{cso09}), one expects the fulfilling of
\begin{equation}\label{cso16}
\frac{d\big<\bm{r\cdot p}\big>}{dt}=0.
\end{equation} 
In order to get a general version of the Virial theorem in terms of complex position and momentum operators, one sets the general position operator to $\widehat{\bm z}$ in (\ref{cso23}) and thus obtains
\begin{equation}\label{cso14}
	\frac{d\big<\bm{z\cdot p}\big>}{dt}=\frac{d\big<\bm{r\cdot p}\big>}{dt}+\left<\bm{s\cdot\nabla}\left(\frac{U-U^\dagger}{2i}\right)\right>+\big<\bm{\sigma\cdot p}\big>,
\end{equation}
where
\begin{equation}\label{cso15}
	\frac{d\big<\bm{r\cdot p}\big>}{dt}=\frac{1}{m}\big<p^2\big>-\big<\bm{r\cdot\nabla}U\big>
\end{equation}
and using the Einstein notation for contracted indices,
\begin{equation}
	\widehat\sigma_k=\frac{i\hbar}{2m}\Big(\nabla^2s_k+2\partial_l s_k\partial^l\Big).
\end{equation}
In principle, the periodicity of the physical system that satisfies (\ref{cso15}) is not equivalent to that encoded in (\ref{cso14}), and  the whole expression may not be zero, because the generalized position operator may involve non-stationary processes. It is nevertheless reasonable to consider (\ref{cso14}) as a generalized version of the Virial theorem for quantum non-stationary processes, and comparative research involving classical mechanics seems to be an interesting direction for future research.

Following an identical method, the generalization of (\ref{cso15}) in terms of the generalized momentum $\widehat{\bm\Pi}$ of real $\bm{\mathcal A}$ is simply
\begin{equation}\label{cso18}
	\frac{d\big<\bm{r\cdot \Pi}\big>}{dt}=\frac{1}{m}\big<\Pi^2\big>-\big<\bm{r\cdot\nabla}U\big>+\frac{q}{mc}\big<\bm{\mathcal B\cdot L}\big>,
\end{equation}
where $\widehat{\bm L}=\widehat{\bm r}\bm\times\widehat{\bm\Pi}$. Therefore, using the full generality of the operator, after some algebra one obtains
\begin{equation}\label{cso17}
	\frac{d\big<\bm{z\cdot \Pi}\big>}{dt}=\frac{1}{m}\big<\Pi^2\big>-\big<\bm{z\cdot\nabla}U\big>+\frac{q}{mc}\big<\bm{\mathcal B\cdot \mathcal L}\big>+\frac{q}{2mc}\big<\bm{\ell\cdot \mathcal B}\big>+\big<\mathcal S\big>,
\end{equation}
where complex angular momentum operators read
\begin{equation}\label{cso19}
	\widehat{\bm\ell}=\widehat{\bm z}\bm\times\widehat{\bm p},\qquad\qquad\widehat{\bm{\mathcal L}}=\widehat{\bm z}\bm\times\widehat{\bm \Pi},
\end{equation}
and the correction term is generated by the operator
\begin{equation}
	\widehat{\mathcal S}=\hbar\left(\nabla^2\bm s\right)\bm{\cdot\Pi}-2i\Big[\hbar^2\partial_k s_\ell\partial^k\partial^\ell+\mathcal A_k\big(\partial^ks_\ell+\partial_\ell s^k\big)\widehat p^{\,\ell}+\big(\partial_ks_\ell\big)\widehat{p}^{\,k}\mathcal A^\ell+\mathcal A_k\mathcal A_\ell\partial^k s^\ell\Big].
\end{equation}
This final result fully generalizes the Virial theorem, and demonstrates the contribution of the imaginary components to the expectation value, as can immediately seen from the $\bm{z\cdot\nabla}U$ contribution to (\ref{cso17}). Although a clearer physical interpretation seems to depend on classical results to be obtained from future research, the result is mainly important because it demonstrates the usual expressions (\ref{cso15}) and (\ref{cso18}) to be generalized in terms of plausible complex operators like (\ref{cso19}) that can be further explored in future research as well.

%%%%%%%%%%%%%%%%%%%%%%%%%%%%%%%%%%%%%%%%%%%%%%%%%%%%%%%%%%%%%%%%%%%%%%%%%%
%%%%%%%%%%%%%%%%%%%%%%%%%%%%%%%%%%%%%%%%%%%%%%%%%%%%%%%%%%%%%%%
\section{LEFT QUATERNIONIC SOLUTIONS\label{lqs}}
%%%%%%%%%%%%%%%%%%%%%%%%%%%%%%%%%%%%%%%%%%%%%%%%%%%%%%%%%%%%%%%
%%%%%%%%%%%%%%%%%%%%%%%%%%%%%%%%%%%%%%%%%%%%%%%%%%%%%%%%%%%%%%%%%%%%%%%%%%

This section and the next are intended to examine the consequences of adopting quaternionic wave functions within the physical cases considered in terms of complex numbers in the previous section. One will not discuss here the mathematical theory of quaternionic numbers that can be found elsewhere  \cite{Ward:1997qcn,Morais:2014rqc,Ebbinghaus:1990zah}, but only remember that a quaternionic number, denoted as $\mathbbm H$, is a non-commutative hyper-complex numeral  written as
\begin{equation}\label{lqs01}
	\Psi=\psi_1+\psi_2 j,
\end{equation}
where $\psi_1$ and $\psi_2$ are complex, and $j$ represents an imaginary unit such that
\begin{equation}\label{lqs03}
i^2=j^2=-1,\qquad\mbox{and}\qquad ij=-ji.
\end{equation}
As a consequence of the anti-commutative property (\ref{lqs03}) involving the imaginary units $i$ and $j$, it follows that
\begin{equation}\label{lqs05}
i\Psi\neq \Psi i,
\end{equation}
whose immediate consequence is the existence of the two possible wave equations to be pondered in the sequel.

\paragraph{LEFT QUATERNIONIC DYNAMICS} The quaternionic non-commutativity allows that quaternionic wave functions $\Psi$ may satisfy two possible quaternionic wave equations, firstly
\begin{equation}\label{lqs02}
	i\hbar\frac{\partial \Psi}{\partial t}=\widehat{\mathcal H}_L\Psi,
\end{equation}
and called left wave equation inspired by the position of the $i$ imaginary unit, and the second possibility  will be entertained into the next section. The left quaternionic Hamiltonian operator contained in (\ref{lqs02}) reads
\begin{equation}
	\widehat{\mathcal H}_L=\frac{1}{2m}\widehat{\mathcal P}_L^2+\mathcal U,
\end{equation}
and comprises the scalar quaternionic potential
\begin{equation}\label{lqs06}
	\mathcal U=U_1+U_2 j
\end{equation}
where $U_1$ and $U_2$ are complex, and the left quaternionic linear momentum operator, namely
\begin{equation}\label{lqs07}
	\widehat{\bm{\mathcal P}}_L=\widehat{\bm p}_L-\frac{q}{c}\bm{\mathrm A},
\end{equation}
comprehending the left linear momentum operator $\widehat{\bm p}_L$, and the quaternionic vector potential $\bm{\mathrm A}$. The action of $\widehat{\bm p}_L$ over a quaternionic function is such as
\begin{equation}\label{lqs09}
	\widehat{\bm p}_L\Psi=-i\hbar\bm\nabla\Psi,
\end{equation}
and the quaternionic vector potential $\bm{\mathrm A}$, unavoidably
\begin{equation}\label{lqs08}
	\bm{\mathrm A}=\bm{\mathcal A}_1+\bm{\mathcal A}_2 j,
\end{equation}
encompasses $\bm{\mathcal A}_1$ and $\bm{\mathcal A}_2$ as complex vector potentials. The real Hilbert space approach also holds for quaternionic wave function and consequently non-hermitian operators are admitted, implying  $\widehat{\bm{\mathcal P}}_L$ and $\widehat{\mathcal H}_L$ to verify (\ref{cso20}) in general, and the expectation value defined within (\ref{cso02}-\ref{cso05}) also applies. Therefore, following the sequence of probes considered in the complex case, one passes to the next topic.

\paragraph{LEFT CONTINUITY EQUATION} Multiplying by $i/\hbar$ the left hand side of the wave equation (\ref{lqs02}), and by $\Psi^\dagger$ the right hand side of the same equation, one obtains the continuity equation after taking the real component, accordingly
\begin{equation}\label{lqs04}
	\frac{\partial \rho}{\partial t}+\Big(\bm\nabla+\bm\Gamma\Big)\bm{\cdot \mathcal J}=\mathsf {g}.
\end{equation}
The probability density $\rho$ and the current of probability density are respectively
\begin{equation}
	\rho=\Psi\Psi^\dagger,\qquad\qquad
	\bm{\mathcal J}=\frac{1}{2m}\left[\big(\widehat{\bm{\mathcal P}}_L\Psi\big)\Psi^\dagger+\Psi\big(\widehat{\bm{\mathcal P}}_L\Psi\big)^\dagger\right].
\end{equation}
and the non-conservative components are
\begin{equation}
	\bm\Gamma=\frac{\bm {i\mathsf A}^\dagger -\bm{\mathsf A}i}{\hbar},\qquad \mathsf g=\frac{\mathcal U^\dagger i-i\,\mathcal U}{\hbar}\varrho.
\end{equation}
The correspondence between the complex case (\ref{cso03}) and (\ref{lqs04}) is exact and impressive, indicating analogous physical interpretations between the cases. 
 
\paragraph{LEFT CLASSICAL LIMIT} One may generalize the complex position operator $\widehat{\bm z}$ defined at (\ref{cso06}) in terms of a quaternionic position operator $\widehat{\bm q}$ postulated as
\begin{equation}\label{lqs16}
	\widehat{\bm q}=\widehat{\bm z}+\widehat{\bm w}\,j,
\end{equation}
where $\widehat{\bm z}$ follows (\ref{cso06}), and
\begin{equation}
	\widehat{\bm w}=\bm w(\bm r).
\end{equation}
By the same token as (\ref{cso21}), the quaternionic position operator satisfies
\begin{equation}
	\big<\bm q\big>=\big<\bm r\big>.
\end{equation}
Moreover, also following the complex case, one obtains
\begin{equation}\label{lqs17}
\Big[\widehat q_\ell,\,\widehat p_{L\,k}\Big]=\hbar \Big(i\delta_{k\ell}-\partial_k s_\ell+\partial_k w_\ell ij\Big)+2w_\ell j\,\widehat p_{L k},
\end{equation}
and (\ref{cso22}) is recovered if $\bm w=\bm 0$, as expected. The investigation of (\ref{cso22}) and (\ref{lqs17}) as alternatives for the generalized uncertainty principle \cite{Lake:2023uoi,Bosso:2023aht} is an exciting direction for forthcoming research.

On the other hand, in analogy to that done in the complex case, one employs the continuity equation (\ref{lqs04}) to obtain
\begin{equation}\label{lqs10}
	\frac{d\big< \bm q\big>}{dt}=\frac{1}{m}\big<\bm{\mathcal P}_L\big>+\big<\bm q\big( \mathsf g-\bm{\Gamma\cdot\mathcal J}\big)\big>,
\end{equation}
and another straightforward calculation permits to reach
\begin{equation}\label{lqs11}
	\frac{d\big< \bm p\big>}{dt}=-\left<\bm\nabla\left(\frac{U+U^\dagger}{2}\right)\right>+\left<\frac{U-U^\dagger}{i\hbar}\,\bm p\right>.
\end{equation}
The similarity between (\ref{lqs10}-\ref{lqs11}) and (\ref{cso07}-\ref{cso08}) is complete, establishing a well-defined classical limit, something that is not achieved within the usual Hilbert space approach to $\mathbbm H$QM ({\em cf.} Sec. 4.4 of \cite{Adler:1995qqm}). The consistency of the classical limit for quaternionic wave functions is likely the most decisive element to endorse the real Hilbert state approach as suitable for generalized quantum theory.

\paragraph{LEFT EXPECTATION VALUES DYNAMICS} It was possible in the complex case to  withdraw the hermitian operator constraint and hence generalize the expectation value dynamics (\ref{cso09}) having the benefit of relation (\ref{cso10}), valid for arbitrary operators.  Using the left quaternionic wave equation (\ref{lqs02}), it is already known from \cite{Giardino:2018lem} that 
\begin{equation}\label{lqs14}
	\Big<\big[\mathcal H_L,\,\mathcal O-i\mathcal O i\big]\Big>=\hbar\left<\frac{\partial}{\partial t}\Big(i\mathcal O+\mathcal O i\Big)\right>.
\end{equation}
Consequently, one obtains
\begin{equation}\label{lqs12}
	\frac{d}{dt}\Big<i\mathcal O+\mathcal O i\Big>=\frac{1}{\hbar}\Big<\big[\mathcal  O-i\mathcal O i,\,\mathcal  H_L\big]\Big>+\left<\frac{\partial}{\partial t}\Big(i\mathcal O+\mathcal O i\Big)\right>+\frac{\partial}{\partial t}\Big<i\mathcal O+\mathcal O i\Big>.
\end{equation}
For practical purposes, it is possible to make $\mathcal O\to i\mathcal O $, in (\ref{lqs12}) to obtain
\begin{equation}\label{lqs13}
	\frac{d}{dt}\Big<\mathcal O-i\mathcal O i\Big>=-\frac{1}{\hbar}\Big<\big[i\mathcal  O+\mathcal O i,\,\mathcal  H_L\big]\Big>+\left<\frac{\partial}{\partial t}\Big(\mathcal O-i\mathcal O i\Big)\right>+\frac{\partial}{\partial t}\Big<\mathcal O-i\mathcal O i\Big>.
\end{equation}
The above result indicates the complex part of the operator as the generator of expectation value of $\mathcal O$, because $\widehat{\mathcal O}-i\widehat{\mathcal O}i$ and $\widehat{\mathcal O}-i\widehat{\mathcal O}i$ are both complex. However, one cannot forget that $\widehat{\mathcal H}_L$ is quaternionic, and eventually sharing physical properties with $\widehat{\mathcal O}$. Therefore, although (\ref{lqs13}) is apparently equivalent to (\ref{cso09}), their physical content differs significantly. 

\paragraph{LEFT QUATERNIONIC LORENTZ FORCE} Choosing 
$\widehat{\mathcal O}=\widehat{\bm{\mathcal P}}_L$ in (\ref{lqs13}), one obtains
\begin{equation}
	\frac{d}{dt}\Big<\bm\Pi_L\Big>=-\frac{1}{\hbar}\left<\Big[i\bm\Pi_L,\,\mathcal  H\Big]\right>+\left<\frac{\partial}{\partial t}\bm\Pi_L\right>+\frac{\partial}{\partial t}\Big<\bm\Pi_L\Big>,
\end{equation}
where $\widehat{\bm\Pi}_L$ is the complex component of $\widehat{\bm{\mathcal P}}_L$. Proceeding in a similar way to that of the complex case, one obtains
\begin{equation}\label{lqs15}
	\frac{d}{dt}\Big<\bm\Pi_L\Big>=\left< \frac{1}{2m\hbar i}\Big( \bm\Pi_L\bm{\times\mathcal B} - \bm{\mathcal B\times\Pi}_L\Big) \right>-\left<\bm\nabla\left( \mathcal U-\frac{q^2}{c^2}\bm{\mathcal A}_2\bm{\cdot\mathcal A}^\dagger_2\right)\right>.
\end{equation}
Seemingly, the quaternionic vector potential affects the expectation value in terms of a contribution to the scalar potential coming from the Hamiltonian operator, and the further terms  conform the complex case (\ref{cso24}), thus identifying it with a generalization of the previous case.

\paragraph{LEFT VIRIAL THEOREM} In this case, using
\begin{equation}
\widehat{\mathcal O}=\widehat{\bm q}\bm\cdot\widehat{\bm p}_L
\end{equation}
one plainly recovers the complex result (\ref{cso14}). On the other hand, from
\begin{equation}
	\widehat{\mathcal O}=\widehat{\bm q}\bm\cdot\widehat{\bm{\mathcal P}}_L
\end{equation}
one obtains
\begin{multline}
\frac{d\big<\bm{q\cdot\mathcal P}_L\big>}{dt}\quad\to\quad\frac{d\big<\bm{z\cdot \Pi}_L\big>}{dt}=\frac{1}{m}\big<\Pi^2\big>-\big<\bm{z\cdot\nabla}\mathcal Q_1\big>+\frac{q}{mc}\big<\bm{\mathcal B\cdot \mathcal L}\big>+\frac{q}{2mc}\big<\bm{\ell\cdot \mathcal B}\big>+\big<\mathcal S\big>-\\-\frac{1}{\hbar}\Big<\mathcal Q_2 \,ij\big(\bm z-\bm z^\dagger\big)\bm\cdot\bm p\Big>-\frac{1}{\hbar}\left<i\Big[\bm{w\cdot \mathcal A}_2,\,p^{\,2}-2\bm{\mathcal A}_1\bm\cdot{\bm p}\Big]\right>,
\end{multline} 
where
\begin{equation}
	\widehat{\mathcal Q}_1=U_1-\frac{q^2}{c^2}\bm{\mathcal A}_2\bm{\cdot\mathcal A}_2^\dagger
\end{equation}
and
\begin{equation}
	\widehat{\mathcal Q}_2=\frac{q^2}{c^2}\bm{\mathcal A}_2\bm\cdot\Big(\bm{\mathcal A}_1+\bm{\mathcal A}^\dagger_1\Big)-\frac{q}{c}\bm p_L\bm{\cdot\mathcal A}_2+U_2.
\end{equation}
Again, this achievement generalizes the complex outcome and reinforces the power of the real Hilbert space approach to deal with the quaternionic case. 

%%%%%%%%%%%%%%%%%%%%%%%%%%%%%%%%%%%%%%%%%%%%%%%%%%%%%%%%%%%%%%%%%%%%%%%%%%
%%%%%%%%%%%%%%%%%%%%%%%%%%%%%%%%%%%%%%%%%%%%%%%%%%%%%%%%%%%%%%%
\section{RIGHT QUATERNIONIC SOLUTION\label{rqs}}
%%%%%%%%%%%%%%%%%%%%%%%%%%%%%%%%%%%%%%%%%%%%%%%%%%%%%%%%%%%%%%%
%%%%%%%%%%%%%%%%%%%%%%%%%%%%%%%%%%%%%%%%%%%%%%%%%%%%%%%%%%%%%%%%%%%%%%%%%%
As in the previous case, one examines the dynamical properties encoded within quaternionic wave functions (\ref{lqs01}). However, the anti-commuting property (\ref{lqs05}) admits a wave equation other than (\ref{lqs02}), and the consequences of this possibility are as follows.

\paragraph{RIGHT QUATERNIONIC DYNAMICS} The wave equation to be considered is
\begin{equation}\label{rqs01}
	\hbar\frac{\partial \Psi}{\partial t}i=\widehat{\mathcal H}_R\Psi,
\end{equation}
where the right Hamiltonian amounts to
\begin{equation}
	\widehat{\mathcal H}_L=\frac{1}{2m}\widehat{\mathcal P}_R^2+\mathcal U.
\end{equation}
The quaternionic scalar potential $\mathcal U$ equates the (\ref{lqs06}) previous definition, and the right quaternionic linear momentum operator is
\begin{equation}\label{rqs05}
	\widehat{\bm{\mathcal P}}_R=\widehat{\bm p}_R-\frac{q}{c}\bm{\mathrm A}.
\end{equation}
The definition of the quaternionic vector potential accompanies (\ref{lqs08}), and the difference of the right case (\ref{rqs05}) from (\ref{lqs07}) resides in the definition of the right linear momentum operator $\widehat{\bm p}_R$, whose action over a quaternionic function $\Psi$ is such as
\begin{equation}
	\widehat{\bm p}_R\Psi=-\hbar\bm\nabla\Psi i,
\end{equation}
whose single difference to the left linear momentum operator (\ref{lqs09}) is the position of the imaginary unit $i$.

\paragraph{RIGHT CONTINUITY EQUATION} After multiplying the wave equation (\ref{rqs01}) by $i\Psi^\dagger$ and taking the real component, one obtains the right continuity equation
\begin{equation}\label{rqs06}
	\frac{\partial \varrho}{\partial t}+\bm{\nabla\cdot \mathcal J}=\mathsf {g}+\gamma,
\end{equation}
whose probability density and probability density current are respectively
\begin{equation}
	\varrho=\Psi\Psi^\dagger,\qquad\qquad
	\bm{\mathcal J}=\frac{1}{2m}\left[\big(\widehat{\bm{\mathcal P}}_R\Psi\big)\Psi^\dagger+\Psi\big(\widehat{\bm{\mathcal P}}_R\Psi\big)^\dagger\right],
\end{equation}
and the non-conservative terms are
\begin{equation}\label{rqs02}
	\gamma=\frac{1}{2m\hbar}\mathfrak{Re}\left[\bm{\mathrm A\cdot}\Big(\big(\widehat{\bm{\mathcal P}}_R\Psi\big)i\Psi^\dagger+\Psi i\big(\widehat{\bm{\mathcal P}}_R\Psi\big)^\dagger\Big)\right],
	\qquad \mathsf g=\frac{1}{\hbar}\Big(\Psi i\Psi^\dagger\mathcal U^\dagger -\mathcal U \Psi i\Psi^\dagger\Big).
\end{equation}
One observes a difference between (\ref{rqs06}) and the continuity equation (\ref{lqs04}) in terms of the definition of the density current, as well as of the definition of the non-conservative terms. This fact indicates to non-equivalent quaternionic dynamics, where the contribution to non-conservative terms have distinct origin in each case. 

\paragraph{RIGHT CLASSICAL LIMIT} 
The introduction of the quaternionic position operator (\ref{lqs16}) determined the commutation relation (\ref{lqs17}), and a counterpart in the right quaternionic case is required. After introducing the notation
\begin{equation}
	(p|q)r=prq,\qquad\forall p,\,q,\,r\in\mathbbm H,
\end{equation}
that one may write illustratively
\begin{equation}
	\widehat{\bm p}_R\Psi=-\hbar\bm\nabla\Psi i=\big(-\hbar\bm\nabla|\,i\big)\Psi,
\end{equation}
one obtains
\begin{equation}
	\Big[\widehat q_\ell,\,\widehat p_{R\,k}\Big]=\hbar \Big(\delta_{k\ell}+\partial_k s_\ell i+\partial_k w_\ell j\,\Big|\, i\Big).
\end{equation}
Again, the difference to the previous (\ref{lqs17}) is evident because there is no dependence on momentum operator, and it is another element to confirm the physical difference between the left and the right quaternionic cases, and the wider range of possibilities ensued by the real Hilbert space approach. 

On the other hand, proceeding like the left quaternionic case (\ref{lqs10}-\ref{lqs11}), one obtains
\begin{equation}
	\frac{d\big< \bm q\big>}{dt}=\frac{1}{m}\big<\bm{\mathcal P}_R\big>+\big<\bm q\big( \mathsf g-\gamma\big)\big>,
\end{equation}
as well as
\begin{equation}\label{rqs03}
	\frac{d\big< \bm p\big>}{dt}=-\left<\bm\nabla\left(\frac{\mathcal U+\mathcal U^\dagger}{2}\right)\right>.
\end{equation}
One observes an interesting difference between the cases, where the imaginary component of the scalar potential does not contribute to (\ref{rqs03}), although it contributes to (\ref{lqs11}), another interesting demonstration of the non-equivalence between the left and right quaternionic cases.

\paragraph{RIGHT EXPECTATION VALUES DYNAMICS} This instance, alike (\ref{lqs14}) and also following \cite{Giardino:2018lem}, one obtains 
\begin{equation}
	\left[\widehat{\mathcal H}_R,\,\widehat{\mathcal O}\right]=\hbar\left(\frac{\partial\widehat{\mathcal O}}{\partial t}\Bigg|\,i\right),
\end{equation}
and thus
\begin{equation}
	\left(\left[\widehat{\mathcal H}_R,\,\widehat{\mathcal O}\right]\Big|\,i\right)=-\hbar \frac{\partial\widehat{\mathcal O}}{\partial t}.
\end{equation}
Subsequently, likewise to (\ref{lqs13}) one accomplishes
\begin{equation}\label{rqs07}
	\frac{d\big<\mathcal O\big>}{dt}=-\left<\frac{1}{\hbar}\Big(\big[\mathcal  O,\,\mathcal  H_R\big]\big|\,i\Big)\right>+\left<\frac{\partial\mathcal O}{\partial t}\right>+\frac{\partial\big<\mathcal O\big>}{\partial t}.
\end{equation}
One observes the quaternionic components of the operator $\widehat{\mathcal O}$ contributing to the dynamical equation and that represents a noticeable difference related to the left quaternionic dynamics (\ref{lqs13}) where only the complex component of the operator $\widehat{\mathcal O}$ contributes to the expectation value,  and thus the physical phenomenology of each case  differs appreciably.

\paragraph{RIGHT QUATERNIONIC LORENTZ FORCE} In this case, one considers
\begin{equation}
	\frac{d\big<\bm{\mathcal P}_R\big>}{dt}=-\left<\frac{1}{\hbar}\left(\big[\bm{\mathcal P}_R,\,\mathcal  H\big]\big|\,i\right)\right>+\left<\frac{\partial\bm{\mathcal P}_R}{\partial t}\right>+\frac{\partial\big<\bm{\mathcal P}_R\big>}{\partial t}.
\end{equation}
Following \cite{Giardino:2018rhs}, where further details of this calculation can be found, one obtains
\begin{equation}
	\frac{1}{\hbar}\Big[\widehat{\mathcal P}_{Ri},\,\widehat{\mathcal P}_{Rj}\Big]\Psi\,i=-\frac{q}{c}\epsilon_{ijk}\mathrm B_k\,\Psi
\end{equation}
where
\begin{equation}
	\widehat{\bm{\mathrm B}}=\Bigg[\,\bm{\nabla\times\mathrm A} -\frac{q}{\hbar \,c}\big(\bm{\mathrm A\times\mathrm A}\big|\,i\big)\Bigg],
\end{equation}
remembering the vector product not to be zero because of the quaternionic quantities involved.
Finally,
\begin{equation}\label{rqs04}
	\frac{d\big<\bm{\mathcal P}_R\big>}{dt}=\frac{1}{\hbar}\left<\frac{\bm{\Pi\times\mathrm B} - \bm{\mathrm B\times\Pi}}{2m}\right>-\big<\bm\nabla\mathcal U\big>.
\end{equation}
Comparing  (\ref{lqs15}) and the left result (\ref{rqs04}), one observes that each case offers different corrections to (\ref{cso13}), the left through the scalar potential, and the second through the vector potential, configuring a further evidence of the different physical content of each quaternionic dynamical equation.

\paragraph{RIGHT VIRIAL THEOREM} In this final section, after choosing
\begin{equation}
	\widehat{\mathcal O}=\widehat{\bm q}\bm\cdot\widehat{\bm p}_R
\end{equation}
one obtains the right quaternionic version of the Virial theorem, whose complete expression comprises
\begin{equation}
	\frac{d\big<\bm{q\cdot p}_R\big>}{dt}=\frac{d\big<\bm{z\cdot p}_R\big>}{dt}-\frac{1}{\hbar}\Big<\Big(\big[\bm{z\cdot p}_R,\,U_2j\big]\Big|i\Big)\Big>-\frac{1}{\hbar}\left<\Big(\Big[\bm w j\bm{\cdot p}_R,\,{\mathcal H}_R\Big]\Big|i\Big)\right>,
\end{equation}
and the quaternionic terms correcting the complex dynamics (\ref{cso14}) accordingly comprehend
\begin{equation}
	\frac{1}{\hbar}\Big<\Big(\big[\bm{z\cdot p}_R,\,U_2j\big]\Big|i\Big)\Big>=\Big<U_2 j\big(\bm z-\bm z^\dagger\big)\bm{\cdot\nabla}\Big>,
\end{equation}
and
\begin{equation}
\frac{1}{\hbar}\left<\Big(\Big[\widehat{\bm w}j\bm\cdot\widehat{\bm p}_R,\,\widehat{\mathcal H}_R\Big]\Big|i\Big)\right>=\frac{\hbar^2}{2m}\Big<\Big(\partial_k\partial^k w_\ell+2\partial_k w_\ell\partial^k\Big)j\partial^\ell\Big>+\Big<\Big(w_\ell j\mathcal U-\mathcal U w_\ell j\Big)\partial^\ell\Big>.
\end{equation}
This result confirms the right quaternionic version of the Virial theorem as a correction to the complex case that is non-equivalent to the left quaternionic case, that is equivalent to the complex result. As a final point, one observes that it is possible to set
\begin{equation}
	\widehat{\mathcal O}=\widehat{\bm q}\bm\cdot\widehat{\bm{\mathcal P}}_R,
\end{equation}
and obtain the precise corrections to the complex case using
\begin{equation}
	\widehat{\bm q}\bm\cdot\widehat{\bm{\mathcal P}}_R=\widehat{\bm z}\bm\cdot\widehat{\bm{\Pi}}_R+\widehat{\mathcal R}
\end{equation}
where
\begin{equation}
\widehat{\mathcal R}=\widehat{\bm w}\bm\cdot\widehat{\bm{\mathcal A}}_R^\dagger-\widehat{\bm z}\bm\cdot\widehat{\bm{\mathcal A}}_2 j+\widehat{\bm w} j\bm\cdot\widehat{\bm{\Pi}}_R.
\end{equation}
The exact result may depend on exact expression of $\bm s$ and $\bm w$, but the important point is the confirmation of the quaternionic version as a generalization of the complex case, and the non-equivalence between the quaternionic proposals. The research of concrete physical situations is an exciting direction for future research.

%%%%%%%%%%%%%%%%%%%%%%%%%%%%%%%%%%%%%%%%%%%%%%%%%%%%%%%%%%%%%%%%%%%%%%%%%%
%%%%%%%%%%%%%%%%%%%%%%%%%%%%%%%%%%%%%%%%%%%%%%%%%%%%%%%%%%%%%%%
\section{CONCLUSION\label{OCO}}
This article examined the expectation value quantum dynamics within the real Hilbert space formalism, enabling the admission of non-hermitian operators among the physically significant operators. The approach also enables the introduction of imaginary components in the scalar potential, as well as in the vector potential, and immediately admitting generalized forms of the position, and the angular momentum operators. This generalization does not affect the position and linear momentum expectation value, and can be understood as gauge transformations whose effect appear in situations where further interactions are allowed.  This was verified after developing the expectation values dynamics, and the application of it to the quantum Lorentz force, as well as to the Virial theorem. The results confirm the generalization of the $\mathbb C$QM results in terms of corrections to expectation values, and also demonstrate the actual value of these corrections to depend on the either complex or quaternionic character of the wave function, and of the operator.

The reported results confirm and extend previous $\mathbbm H$QM  study \cite{Giardino:2018rhs}, and one stresses the generalized position and angular momentum operators, the uncertainty relation obtained from them, the expectation value dynamics, as well as the left quaternionic wave function results as clear novelties presented here. Future aspects to be investigated may include angular momentum and spin theory. However, the real Hilbert space has further application in almost every area of quantum mechanics, and one quotes perturbation theory, and adiabatic theories as the most immediate possible applications to be investigated.

%%%%%%%%%%%%%%%%%%%%%%%%%%%%%%%%%%%%%%%%%%%%%%%%%%%%%%%%%%%%%%%
%%%%%%%%%%%%%%%%%%%%%%%%%%%%%%%%%%%%%%%%%%%%%%%%%%%%%%%%%%%%%%%%%%%%%%%%%%

\begin{footnotesize}
\paragraph{Funding} The author gratefully thanks for the financial support by Fapergs under the grant 23/2551-0000935-8 within Edital 14/2022.

\paragraph{Data availability statement} The author declares that data sharing is not applicable to this article as no data sets were generated or analyzed during the current study.

\paragraph{Declaration of interest statement} The author declares that he has no known competing financial interests or personal relationships that
could have appeared to influence the work reported in this paper.
\end{footnotesize}
%%%%%%%%%%%%%%%%%%%%%%
%
% 
%  BIBLIOGRAPHY
%
%

%\begin{footnotesize}
%\bibliographystyle{unsrt} 
%\bibliography{bib_lorentz}
%\end{footnotesize}
\end{document}